\documentclass[%
 reprint,
unsortedaddress,
 amsmath,amssymb,
 aps
]{revtex4-2}

\usepackage{graphicx}
\usepackage{dcolumn}
\usepackage{bm}
\usepackage{multirow}
\usepackage{xcolor}
\usepackage{CJKutf8}
\DeclareUnicodeCharacter{2212}{-}

\begin{document}

\begin{CJK*}{UTF8}{bsmi}
\preprint{APS/123-QED}

\title{Signatures of the charge density wave collective mode in the infrared optical response of VSe$_{2}$}

\author{Xuanbo Feng (馮翾博)}
\affiliation{QuSoft, Science Park 123, 1098 XG Amsterdam, The Netherlands}
\affiliation{Institute of Physics, University of Amsterdam, Science park 904, 1098 XH Amsterdam, The Netherlands}
\author{Jans Henke}
\author{Corentin Morice}
\affiliation{Institute of Physics, University of Amsterdam, Science park 904, 1098 XH Amsterdam, The Netherlands}
\author{Charles J. Sayers}
\affiliation{Dipartimento di Fisica, Politecnico di Milano, 20133 Milano, Italy}
\affiliation{Centre for Nanoscience and Nanotechnology, Department of Physics, University of Bath, Bath BA2 7AY, United Kingdom}
\author{Enrico Da Como}
\affiliation{Centre for Nanoscience and Nanotechnology, Department of Physics, University of Bath, Bath BA2 7AY, United Kingdom}
\author{Jasper van Wezel}
\affiliation{Institute of Physics, University of Amsterdam, Science park 904, 1098 XH Amsterdam, The Netherlands}
\author{Erik van Heumen}
\email{e.vanheumen@uva.nl}
\affiliation{QuSoft, Science Park 123, 1098 XG Amsterdam, The Netherlands}
\affiliation{Institute of Physics, University of Amsterdam, Science park 904, 1098 XH Amsterdam, The Netherlands}

\date{\today}

\begin{abstract}
We present a detailed study of the bulk electronic structure of high quality VSe$_{2}$ single crystals using optical spectroscopy. Upon entering the charge density wave phase below the critical temperature of 112 K, the optical conductivity of VSe$_{2}$ undergoes a significant rearrangement. A Drude response present above the critical temperature is suppressed while a new interband transition appears around 0.07\,eV. From our analysis, we estimate that part of the spectral weight of the Drude response is transferred to a collective mode of the CDW phase. The remaining normal state charge dynamics appears to become strongly damped by interactions with the lattice as evidenced by a mass enhancement factor $m^{*}/m\approx$\,3. In addition to the changes taking place in the electronic structure, we observe the emergence of infrared active phonons below the critical temperature associated with the $4a\times 4a$ lattice reconstruction. 
\end{abstract}

\maketitle
\end{CJK*}

\section{Introduction}
Early studies of the transition-metal dichalcogenide (TMDC) VSe$_2$ focussed on the charge density wave transition \cite{Bayard1976, VanBruggen1976,Sugai1981, Tsutsumi1982, Bayliss1984, claessen_JPCM_1990}. The recently renewed interest in TMDCs as a platform for 2D materials research has also rekindled the interest in VSe$_{2}$ \cite{Terashima2003, Sato2004, Strocov2012, Pasztor2017, Barua:SRep2017, Jolie2019}, and in particular in mono-atomic layers of VSe$_{2}$ \cite{bonilla_NatMat_2018, chen_PRL_2018}.  Nevertheless, the characteristics of the CDW phase in the bulk compound, such as the collective excitations and size of the CDW gap, remain incomplete and controversial. For example, angle resolved photoemission (ARPES) and scanning tunnelling spectroscopy (STM) experiments report gap sizes varying between 13 to 130\,meV \cite{Terashima2003, Sato2004, Ekvall1999,Wang1991, Jolie2019}.
Recently, some of us reported the sensitivity of the CDW properties to defects and overall stoichiometry \cite{Sayers2020}. It was demonstrated that depending on sample growth conditions the defects and Se deficiencies proliferate, while optimal growth conditions can be achieved to produce stoichiometric, clean crystals with an enhanced critical transition temperature for the CDW phase. This development opens the door to new studies of bulk electronic properties and here we report the first infrared optical spectroscopy study of the CDW transition of VSe$_{2}$. In the following we present the absolute reflectivity of high-quality VSe$_{2}$ single crystals, and discuss the evolution of the main features across the CDW transition. Our optical data indicates that a significant reorganisation of the electronic structure takes place and we discuss how this compares to earlier experiments that probe the electronic structure. 

\section{Experiment}
The preparation and characterisation of single crystals of VSe$_{2}$ are extensively reported in Ref.\,\cite{Sayers2020}. In summary, it was reported that in order to obtain near ideal stoichiometry and to minimise the presence of defects, samples were prepared by chemical vapour transport at a growth temperature of 550 $^{\circ} C$. Electrical transport measurements show a residual resistance ratio or RRR = 49 and a CDW transition temperature, $T_{c}$ = 112.7 K. The work reported here was carried out on a single crystal with approximate dimensions of 2 $\times$ 3\,mm and an approximate thickness of $70\pm 30\,\mu m$. Reflectivity measurements were performed over the energy range 6\,meV to 4\,eV using a VERTEX 80v FTIR spectrometer as described in Ref.\,\cite{Tytarenko2015}. To obtain the temperature dependent reflectivity, the measurements were designed to run through cooling and warming cycles between 14\,K and 300\,K. In each cycle we used heating or cooling rates of 3.33\,K/min., while an infrared spectrum was recorded every minute. Such temperature cycles were repeated two to three times to increase the signal-to-noise ratio and ensure reproducibility. In order to obtain the absolute reflectivity, all measurements were repeated on 
references obtained by \textit{in-situ} evaporation of reference materials (e.g. Au, Ag or Al). To obtain the full measured energy range we used various detectors, sources and beamsplitters. The absolute reflectivity data were analysed using RefFit software to obtain the optical response functions \cite{Kuzmenko2005}. The procedure starts with the development of a Drude-Lorentz model that accurately describes the measured reflectivity data. This model is used together with the variational dielectric function method described in Ref.\,\cite{Kuzmenko2005} to effectively perform the Kramers-Kronig transformation of the reflectivity data. 

\section{Results}

\begin{figure}
    \includegraphics[width=0.95\columnwidth]{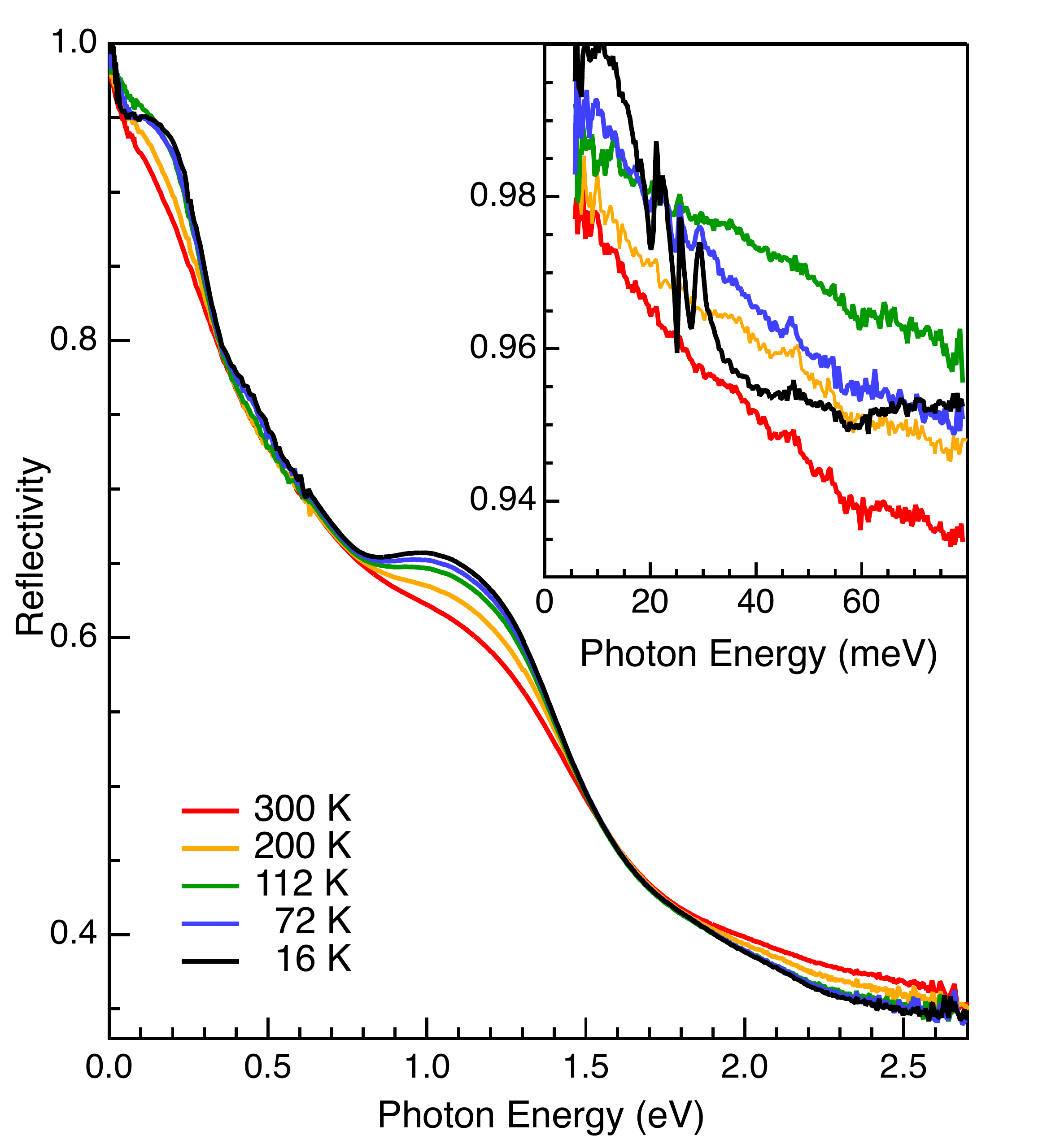}
    \caption{(Colour online) Reflectivity data of VSe$_{2}$ at selected temperatures. The inset presents the far-infrared range of the reflectivity, where optical phonons and other CDW related features can be clearly observed.}
    \label{fig:r}
\end{figure}

Fig.\,\ref{fig:r} presents the reflectivity of VSe$_{2}$ over a broad energy range, while the inset highlights the low energy range. From transport measurements \cite{Bayard1976, VanBruggen1976, Pasztor2017, Barua:SRep2017, Sayers2020} we know that VSe$_2$ is metallic in the normal state and although the temperature dependence changes close to the CDW transition, the resistivity remains metallic down to the lowest temperatures. Our reflectivity data is also indicative of a metallic response, as all data extrapolates to unity at zero frequency. In the normal state we can use the Drude model for metals to obtain a simple approximation for the low frequency reflectivity, also known as the Hagen-Rubens relation, to estimate the DC conductivity.

In the normal state, the reflectivity data of VSe$_2$ below 25\,meV agrees well with the Hagen-Rubens approximation (see Appendix\,\ref{app:hr}). In the CDW state, the experimental data no longer follows a purely square root frequency dependence, while it still approximately extrapolates to unity. This becomes even more prominent for T\,$\leq$\,32\,K, where the reflectivity becomes frequency independent below 10\,meV. This behaviour resembles a reststrahlen band or optical band gap similar to what is seen in superconductors below the critical temperature (see e.g. Ref. \cite{Gruner1988}). In s-wave superconductors with a full gap opening around the Fermi surface, all optical transitions below the gap are suppressed. Consequently, photons impinging on the surface with energy lower than $2\Delta$ are fully reflected. If this interpretation would be correct we would estimate that the optical gap is of order 2$\Delta\approx$\,10\,meV, which is still two times smaller than the lowest value reported by scanning tunnelling microscopy\cite{Jolie2019}. This value is also rather small assuming a BCS type relation between the gap and $T_{c}$. Finally, we also observe a set of four sharp features which emerge below $T_{c}$ and that are clearly visible in the 16\,K data presented in Fig. \ref{fig:r}. These modes are undetected in the normal state and gain prominence with decreasing temperature. We interpret these modes as optical phonons that become infrared active in the CDW phase. The energies of these peaks are temperature independent within our experimental resolution and are summarized in Table \ref{tab:table3} in Appendix \ref{app:dl}.

Fig.\,\ref{fig:s1}a presents the real part of the optical conductivity $\sigma_{1}(\omega)$ obtained from the reflectivity data using the variational dielectric function approach \cite{Kuzmenko2005}. For the extrapolations outside our experimental window we have used the Drude-Lorentz model as described in Appendix\,\ref{app:dl}. Based on our modelling we observe at least five interband transitions, centred around 0.07\,eV, 0.6\,eV, 1.1\,eV, 2\,eV and 2.6\,eV. The lowest energy transition is only observed in the CDW phase, which could indicate that it emerges from new interband transitions associated with band folding in the CDW phase. We note that the transitions around 0.07\,eV and 1\,eV show a strong temperature dependence, while the other transitions do not. 
\begin{figure}
\includegraphics[width=0.95\columnwidth]{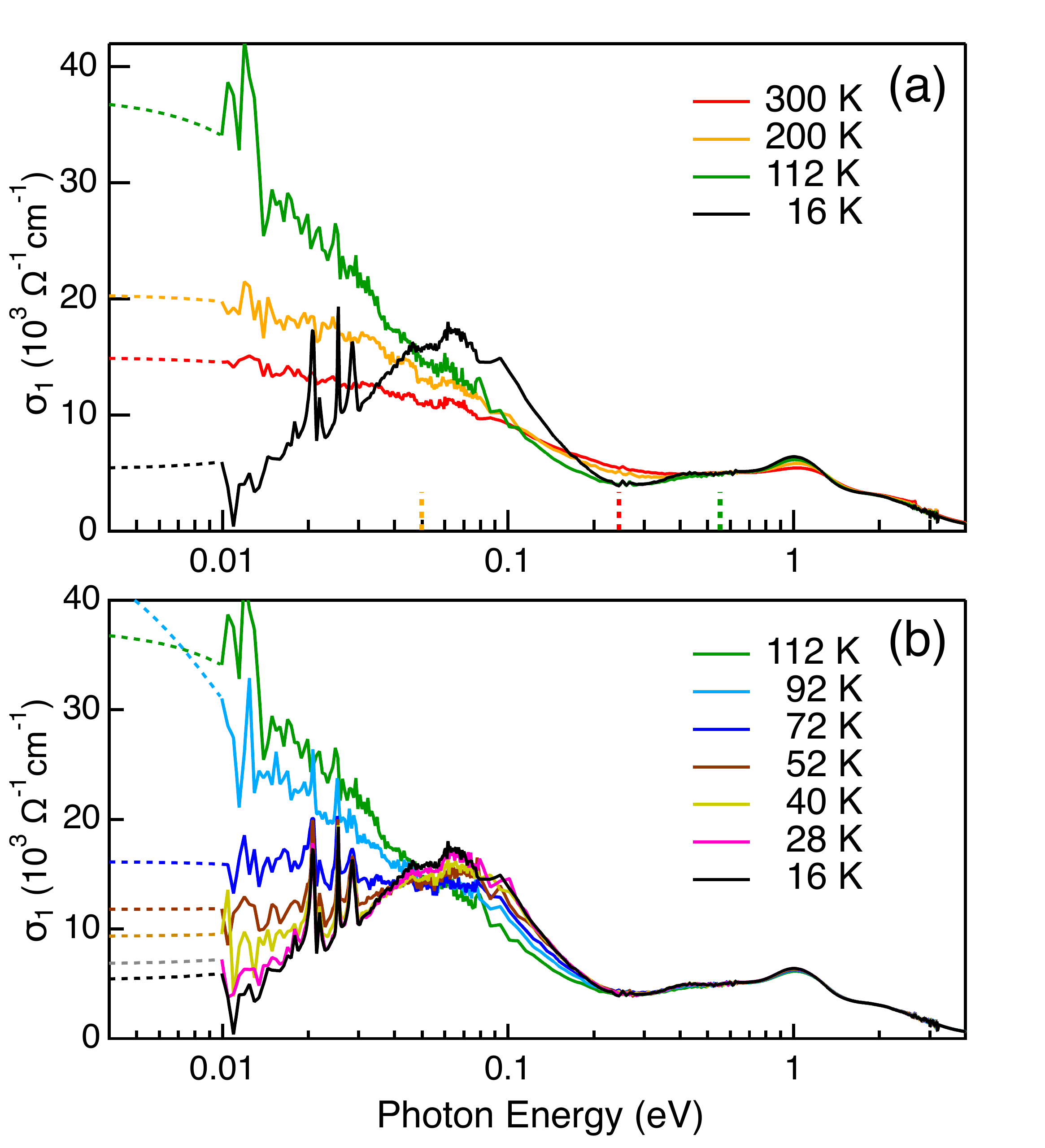}
\caption{(Colour online) (a): Real part of the optical conductivity on logarithmic energy scale at selected temperatures. Dashed lines are extrapolations to low energy derived from the Drude Lorentz models. At room temperature a clear Drude response dominates the low energy range. In contrast, the 16\,K data appears to be gapped with a possible remnant of a Drude peak below our experimental window. The short, vertical dashed lines indicate the cut-off energies for the integrated spectral weight curves presented in Fig. \ref{fig:sw}. (b): similar, but with temperatures spanning the range between the lowest measured temperature and T$_{c}$.}
\label{fig:s1}
\end{figure}

The evolution of the optical conductivity in the CDW phase is presented in Fig. \ref{fig:s1}b. Below 0.2\,eV, the optical conductivity is dominated by a Drude peak at room temperature, which becomes narrower as the temperature is reduced. At temperatures below the CDW transition temperature, the far-infrared optical response undergoes a strong suppression and this removal of spectral weight goes hand in hand with an enhancement of the optical transition centred around 0.07\,eV. The modelling of the reflectivity indicates that part of the spectral weight remains as a broad Drude response with approximately half of the total low spectral weight below 50 meV. This broad Drude response is visible in panel 2b as the flat conductivity indicated by dashed lines (below our experimental window). 

The Drude-Lorentz modelling further indicates that the remainder of the spectral weight moves to a very sharp mode with a lifetime broadening well below our experimental window, that is necessary to reproduce the reflectivity plateau below 10 meV. 
An important aspect of CDW phases is that the breaking of translational symmetry should give rise to a Goldstone mode \cite{Gruner1988}. In the context of CDW phases, these are better known as sliding modes, where the charge density wave modulation moves freely against the periodicity of the underlying lattice. Such sliding modes contribute to the real part of the optical conductivity at zero frequency. However, when the periodicity of the CDW modulation is commensurate with the lattice periodicity, the sliding modes gets pinned to the underlying lattice and will require a finite energy to slide. This pushes the contribution in the optical response to finite frequency and this could be the source for the sharp mode in our Drude-Lorentz model. 

\begin{figure}
\includegraphics[width=0.95\columnwidth]{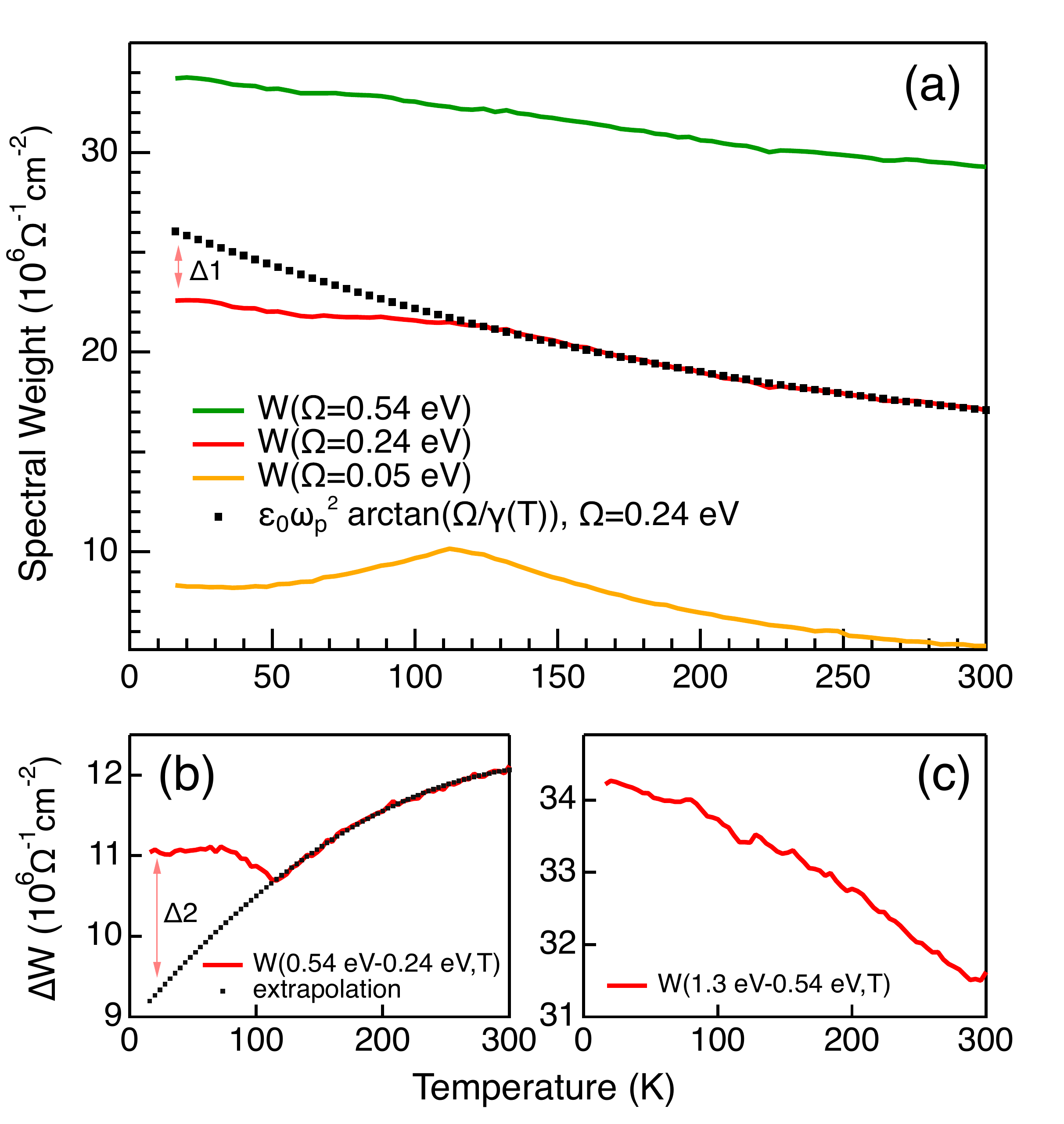}
\caption{(Colour online) (a): Integrated spectral weight of the optical conductivity for selected cut-off energies. Shown are integrals from 0 to 0.05\,eV (gap energy scale), 0.24\,eV (covering both the gap and coherence energy range) and to 0.54\,eV (including the first interband transition). The dotted line presents the fit of the normal state spectral weight using Eq.\,\ref{eq:swfit}. Also indicated is the difference, $\Delta 1$, which is the missing spectral weight in the CDW phase (see text for details). (b) Spectral weight contained in the mid-infrared peak between 0.24\,eV and 0.54\,eV. At $T_{c}$ this peak gains spectral weight, $\Delta 2$, as can be seen by comparing data to to the extrapolated normal state trend (indicated by the dashed line). (c) Partial spectral weight integral for the 1\,eV interband transition. No notable change in the spectral weight temperature dependence is observed around $T_{c}$.}
\label{fig:sw}
\end{figure}

Importantly, these modes are optically active and their formation can be observed from a careful analysis of the optical sum rules \cite{Lee1974}. An estimate of the sliding mode contribution is obtained from an analysis of the optical spectral weight using the f-sum rule. This rule states that the integral of the optical conductivity, a.k.a. spectral weight, is proportional to the ratio of the charge density to the effective mass \cite{VanderMarel2016}. In Fig.\,\ref{fig:sw}a we present the temperature dependent spectral weight for suitably chosen cut-off frequencies of the optical conductivity integral (indicated by vertical, dashed lines in Fig.\,\ref{fig:s1}). Our choices are determined by the optical conductivity data of Fig.\,\ref{fig:s1}: we present one integral with an energy cutoff at 0.05\,eV (yellow) covering most of the Drude response, but not the peak centred around 0.07\,eV; one integral with a cut-off energy of 0.24\,eV (red) that includes this transition and finally one integral with a cut-off around 0.54\,eV (green) that includes the interband transitions at higher energy. We observe that the spectral weight with low energy cut-off (orange) drops sharply below $T_{c}$. This can be explained as a sudden depletion of spectral weight that sets in at $T_{c}$, which is linked to the suppression of the Drude peak. 

The red curve presented in Fig.\,\ref{fig:sw}a integrates the data to 0.24\,eV, which is the energy of the valley in the low temperature optical conductivity data. If the CDW transition simply involves a transfer of spectral weight from intraband to interband response (for example, by opening a gap around the entire Fermi surface), we expect that the spectral weight is transferred to the prominent, optical transition centered at 0.07 eV, which appears below $T_{c}$. We see that there is still a noticeable change of slope taking place at $T_{c}$ in the red curve, indicating that spectral weight is transferred outside this energy window. Finally, we plot the spectral weight integrated up to 0.54\,eV (shown in green), which includes the interband transition that is already visible in the 300\,K data. Here, the change of slope at $T_{c}$ is difficult to discern, but a more careful analysis will show that the spectral weight is still not fully recovered. 

Apart from spectral weight transfer between intraband and interband transitions, one also needs to consider that the collective excitations of the CDW phase could respond to applied electromagnetic fields. The energy scale where these modes typically appear is well below our experimental window, so we expect that some of the missing spectral weight is transferred to the energy range below our measured data ($\omega<$\,10\,meV). To estimate the spectral weight of such a collective excitation we make use of the Ferrell-Glover-Tinkham (FGT) sum rule \cite{Tinkham1959, Ferrell1958}. This sum rule states that the difference between normal state spectral weight and CDW state spectral weight is proportional to the spectral weight of the collective mode, W$_{CM}$: 
\begin{equation}
	W_
	{CM}= \int_{0+}^{\Omega} \left[\sigma_{1,N}(\omega,T)-\sigma_{1,CDW}(\omega,T)\right]\, \mathrm{d}\omega~.
	\label{eq:fgt}
\end{equation}
where $\sigma_{1,N}(\omega,T)$ is the normal state optical conductivity and $\sigma_{1,CDW}(\omega,T)$ the optical conductivity of the CDW state and $\Omega$ is the cut-off energy. Note that the integral starts at a lower cutoff energy $0+$, indicating that the $\omega=0$ contribution is not included. The difficulty with this sum rule is that the conductivities have to be compared at the same temperature. Since the low temperature, normal state optical conductivity is not accessible, one typically has to resort to extrapolations of the normal state spectral weight.

Returning to Fig.\,\ref{fig:sw}, we observe a strong temperature dependence in the normal state spectral weight integrated up to 0.24\,eV. Different interpretations for the temperature dependence of the integrated spectral weight with finite cut-off energy have been considered in the context of the cuprate high-$T_{c}$ superconductors \cite{Benfatto2006, Karakozov2006, Marsiglio2008}. In Ref. \cite{Marsiglio2008} it is shown that a relatively small cut-off energy of the spectral weight integral (compared with the scattering rate $\gamma$) and the sharpening of the Drude response with decreasing temperature results in a transfer of spectral weight from high to lower energies. The size of this effect can be modelled starting from the simple Drude model. We integrate the Drude conductivity to find: 
\begin{equation}
\int_0^\Omega \mathrm{d}\omega~\sigma_{1,D}(\omega) =\varepsilon_0~\omega_p^2\,\arctan\left(\frac{\Omega}{\gamma(T)}\right)~,
\label{eq:swfit}
\end{equation}
in which $\varepsilon_0$ is the vacuum permittivity, $\omega_p$ is the plasma frequency (assumed to be temperature independent), $\gamma(T)$ is the temperature dependent scattering rate and $\Omega$ is the cut-off energy \cite{Marsiglio2008}. From this it is straightforward to show that for $\Omega \gg \gamma$, the Drude weight is temperature independent. However, when the cutoff energies are of order of the scattering rate or smaller, Eq.\,\ref{eq:swfit} attains a temperature dependence. Fermi liquid theory predicts $\gamma(T)=\gamma_0+\beta\,T+\alpha\,T^2$ \cite{Ashcroft76} for the scattering rate where the three terms come from impurity scattering, electron-phonon coupling and electron-electron interactions respectively. Together with the plasma frequency as free parameter, we can use this to fit the normal state temperature dependence and extrapolate the temperature dependence of the normal state spectral weight to zero temperature. The result for cutoff energy $\Omega$\,=\,0.24\,eV is shown as the dotted line in Fig.\,\ref{fig:sw}. From this extrapolation and the measured spectral weight we can estimate the difference at 16\,K (indicated as $\Delta 1$). This gives an estimate of the missing spectral weight, $\Delta W$\,=\,$4\cdot10^6\,\mathrm{\Omega^{−1}cm^{-2}}$. 

One possibility is that this spectral weight goes to even higher energy (into the range up to 0.5 eV). Unfortunately, it is not possible to use the same extrapolation (based on Eq. \ref{eq:swfit}) for higher cutoff energies, due to the presence of additional interband transitions. Instead, we can estimate how much spectral weight is transferred to the high energy range by calculating the spectral weight integral from a lower to an upper bound, e.g. from 0.24\,eV to 0.54\,eV. The temperature dependence of this integral, $\Delta W$(0.54\,-0.24\,eV,T) is shown in Fig.\,\ref{fig:sw}b and shows that indeed some additional spectral weight accumulates in this range as is evidenced by the sudden upturn at T$_{c}$. We use a simple parabolic temperature dependence to extrapolate the normal state temperature dependence and estimate the difference between the measured data and the extrapolation, indicated by $\Delta 2$, as $\Delta 2$\,=\,-\,$2\cdot10^6\,\mathrm{\Omega^{−1}cm^{-2}}$. To exclude that the remaining spectral weight is transferred to even higher energy, Fig.\,\ref{fig:sw}c shows the integrated spectral weight $\Delta W$(1.3\,-0.54\,eV,T). In this energy range we do not observe any CDW related changes. We are now in a position to determine the missing spectral weight according to the FGT sumrule, Eq. \ref{eq:swfit} as:
\begin{equation}
W_{CM}=\Delta_{1}+\Delta_{2}=2\cdot10^6\Omega^{−1}cm^{-2}
\end{equation}
Summarizing, we find that a significant portion of the spectral weight moves to the interband transition around 0.4\,eV. This leaves  spectral weight missing at finite frequency, which most likely contributes to a collective mode below our experimentally accessible energy range.

\begin{figure}
	\includegraphics[width=0.95\columnwidth]{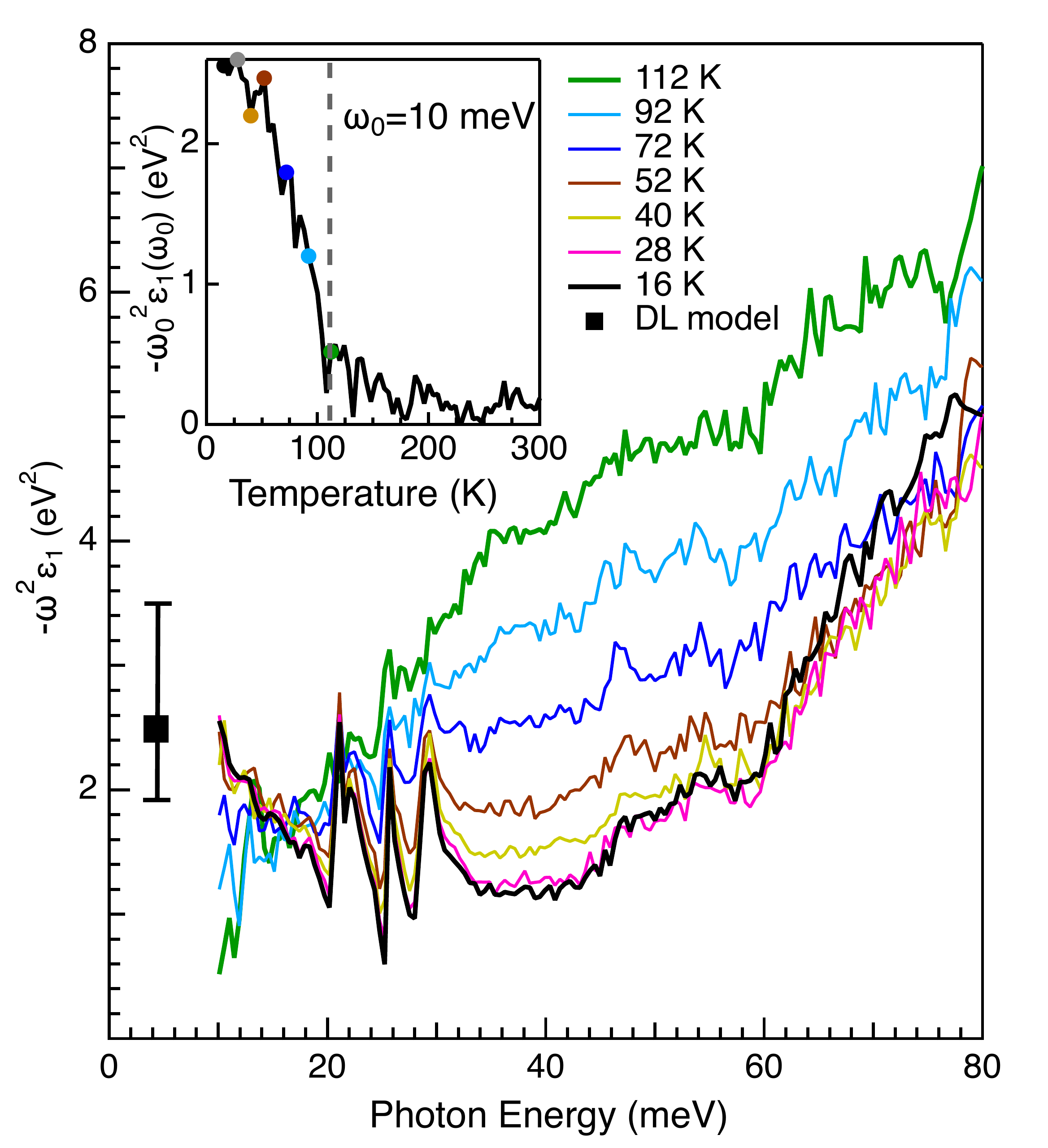}
	\caption{(Colour online) Estimate of the collective mode contribution according to Eq.\,\ref{eq:cm}. The normal state data extrapolates to zero for $\omega\rightarrow 0$. This trend sets in at the critical temperature, as the inset demonstrates by plotting the temperature dependence of the function $-\omega^{2}\varepsilon_{1}(\omega)$ for $\omega_{0}$\,=\,10\,meV. The extrapolation to zero frequency is indicated by the black error bar. The full, black symbol is the value obtained from the Drude-Lorentz model presented in Appendix \ref{app:dl} for comparison.}
	\label{fig:E1}
\end{figure}

An independent test of the presence of a collective mode makes use of the real part of the dielectric function of VSe$_2$, or similarly of the imaginary part of the optical conductivity (see Appendix \ref{app:imsig}). The optical response of the collective mode in the CDW phase is described as a $\delta$-function contribution to $\sigma_{1,CDW}(\omega,T)$ with strength $A(T)\,\delta(\omega-\Omega_0)$ where $\Omega_0$ is the pinning energy of the collective mode and $A(T)$ is a measure of the density of electrons contributing to the CDW phase \cite{Lee1974, Gruner1988}. Both the presence of impurities and coupling to the lattice could result in the pinning of the sliding mode, which moves the $\delta$-function away from zero energy and leads to broadening of the $\delta$-function response. Typical pinning frequencies and broadening factors are expected to be very small compared to the energy scale in our experiment and for our purposes we can consider this as a $\delta$-function response. It is well known that this singular response contributes to the real part of the dielectric function through the Kramers-Kronig relations according to: 
\begin{equation}
	\varepsilon_{1,CDW}=-\frac{A(T)}{\omega^2}
	\label{eq:cm}
\end{equation}
Eq.\,\ref{eq:cm} diverges when $\omega$ approaches zero energy, contrary to the Drude-Lorentz response. This specific signature of the collective mode in the dielectric function can be used to detect the collective mode contribution in the low frequency behaviour of $\varepsilon_1$. By multiplying $\varepsilon_1$ by $\omega^2$, the Drude and Lorentz terms converge to 0 when $\omega\rightarrow0$, while the collective mode response converges to $A(T)$ (see also App. \ref{app:imsig}). This is borne out by the data presented in Fig.\,\ref{fig:E1}, which show that indeed $-\omega^{2}\varepsilon_{1}(\omega,T)$ extrapolates to zero at zero energy in the normal state, while it extrapolates to a finite value at 16\,K. The inset of Fig.\,\ref{fig:E1} shows the temperature dependence of $-\omega^{2}\,\varepsilon_1$ at $\omega_{0}=10\mathrm{meV}$. Below $T_{c}$ this function starts to deviate from zero and suggests the emergence of a 
collective mode resembling Eq.\,\ref{eq:cm} in the CDW phase. 

More importantly, the weight of the collective mode, $A$(T\,=\,16 K) can be extracted by extrapolating to $\omega=0$. The extrapolated range of values are indicated by a black error bar in Fig. \ref{fig:E1}, offset from zero to finite frequency for clarity. This gives an estimate of the spectral weight associated with the collective mode of $4.4\pm 1.5 \cdot10^6\,\mathrm{\Omega^{−1}cm^{-2}}$ with a large uncertainty as a result of our limited low energy data range. The value for the collective mode spectral weight has a range covering our previous estimate, obtained from the spectral weight analysis of the optical conductivity data. Finally, we note that these values also agree with the weight in a collective mode obtained through optimising a Drude - Lorentz model fitted to the reflectivity data (see Appendix \ref{app:dl}). The value obtained from the Drude - Lorentz model is indicated in Fig. \ref{fig:E1} as a black square.

\section{Discussion}
We have found that the optical conductivity undergoes a large reorganisation of the low energy, bulk electronic structure below the charge density wave transition. The Drude peak is suppressed and that this is accompanied by the emergence of a collective mode below our experimental window and an optical transition centred at 0.07\,eV. Next to these observations, one feature remains difficult to reconcile with earlier experiments: the almost complete suppression of the Drude response. The simplest interpretation for this would be the opening of a full gap around the Fermi surface. However, angle resolved photoemission spectroscopy, electrical transport and theoretical studies of VSe$_2$ show that only a small portion of the Fermi surface is gapped in the low temperature phase \cite{Sato2004, Strocov2012, Henke2020}.

We see three possible scenarios for this discrepancy: (i) the matrix elements contributing to the conductivity anomalously enhance the portion of the Fermi surface where a gap opens in the CDW phase, (ii) a strong electron-phonon interaction in the CDW state pushes spectral weight to finite frequency, e.g. a shake-off band or polaron formation or (iii) the temperature dependence is partially governed by band structure effects unrelated to the formation of the CDW state. In the following, we discuss each of these possibilities in turn. 

Within the Kubo formalism, the optical response is governed by dipole matrix elements, $\left<\psi_{f}\right|\hat{v}\left|\psi_{i}\right>$ that weigh the contributions from different parts of the Fermi surface. It is in principle possible that the optical response is anomalously enhanced by the parts of the Fermi surface that are gapped as the CDW phase evolves. From previous work we know that the gap opens only on a small part of the Fermi surface close to the new (folded) zone boundary \cite{Henke2020}. An explicit calculation of the matrix elements (see Appendix \ref{app.me}) shows that the dipole moment associated with the region where the gap opens varies over k-space, but is not significantly larger in the regions where gaps open. 

The second possibility to consider is that the free carrier response is significantly modified by interactions. The spectral weight in the free charge response is proportional to the ratio of carrier density and effective mass according to $\omega_{p}^{2}\propto n/m_{eff}$. It is possible that the formation of the CDW phase is accompanied by a significant mass renormalisation rather than a large change in carrier density. Although it is hard to completely rule out this scenario, we note that dynamical mean field theory calculations of the optical response predict an \textit{enhancement} of spectral weight below the critical temperature for the case of strong electron-phonon coupling \cite{ciuchi_PRB_2008}. Taking the Drude-Lorentz model describing the reflectivity data at face value, we can estimate that the mass enhancement factor associated with the sliding mode is of the order m$^{*}$/m$\approx$\,3. This value is obtained by taking the ratio of the normal state spectral weight of the first Drude mode at $T_{c}$ to the spectral weight in the collective mode \cite{Lee1974}. 

The third possibility is a temperature dependent band structure effect unrelated to the CDW transition itself. Early measurements of the Hall coefficient provide a first clue: it displays a significant temperature dependence already in the normal state indicating a strongly temperature dependent carrier density \cite{VanBruggen1976, Bayard1976,toriumi1981}. Upon entering the CDW phase this trend is even further enhanced. Appendix \ref{app:hr} shows the Hall resistivity measured on a similar crystal used in this study. From the Hall coefficient it follows that the carrier density decreases by an order of magnitude between 220\,K ($n_{H}$\,=\,1.2$\cdot 10^{22}$\,cm$^{-3}$) and 5\,K ($n_{H}$\,=\,0.65$\cdot 10^{21}$\,cm$^{-3}$). Given that the partial gap opening at the Fermi surface cannot be responsible for the large temperature dependent change in the Hall coefficient, we need an alternative explanation. One explanation could be the gradual freezing out of carriers with decreasing temperature associated with band edges close to the Fermi level. Indeed, ARPES data shows a large pocket grazing, but not crossing, the Fermi level around the $\Gamma$ point of the Brillouin zone \cite{Terashima2003, Sato2004,Strocov2012}. At elevated temperatures, thermal excitation of carriers will contribute to both the Hall coefficient and the Drude weight. This contribution freezes out when temperatures becomes smaller than the energy difference between the band maximum and E$_{F}$ (so-called Pauli blocking), resulting in a significant reduction of the carrier density. This scenario could explain the strong reduction of the Drude response we observe in our experiments, as well as the decrease in resistivity due to the removal of an additional scattering channel.

Returning to the spectral weight analysis, we can estimate the relative importance of each of these possibilities explaining the absence of a visible Drude response. The spectral weight associated with the normal state free charge carrier response at $T_{c}$ corresponds to approximately $n_{f}$\,=\,5.8\,-\,9.2$\cdot 10^{21}$\,cm$^{-3}$. These numbers are determined by taking the plasma frequency of the narrow Drude mode reported in Appendix \ref{app:dl} as a lower bound and the combined plasma frequency of the two Drude terms as an upper bound. Depending on how we count the contribution of the two Drude terms, the real value will be somewhere in between. This is consistent with the Hall data provided we take the lower side of this range. At the lowest temperature in our experiments, we find $n_{f}$\,=\,1.6$\cdot 10^{21}$\,cm$^{-3}$ for the normal charge carriers contributing to the optical conductivity. This also agrees with the Hall coefficient data. One interpretation that is consistent with our Drude\,-\,Lorentz model, is that there are additional charge carriers (present at all temperatures) that contribute to transport with a relatively large scattering rate. Such a large scattering rate could well be a result of a relatively strong interaction of these carriers with the lattice, as evidenced by the mass enhancement factor estimated above. This Drude response corresponds to the background conductivity indicated with dashed lines in Fig. \ref{fig:s1}b.

To summarize the preceding discussion, the large suppression of the Drude response most likely results from a convolution of the freezing out of carriers at low temperature and interactions of the remaining electrons with the lattice. In addition, we find strong evidence for the emergence of a sliding mode linked to the formation of the CDW phase in VSe$_{2}$. The remaining feature in our reflectivity data, the plateau in the reflectivity most likely results from this pinned collective mode as we now explain.
\begin{figure}
	\includegraphics[width=0.95\columnwidth]{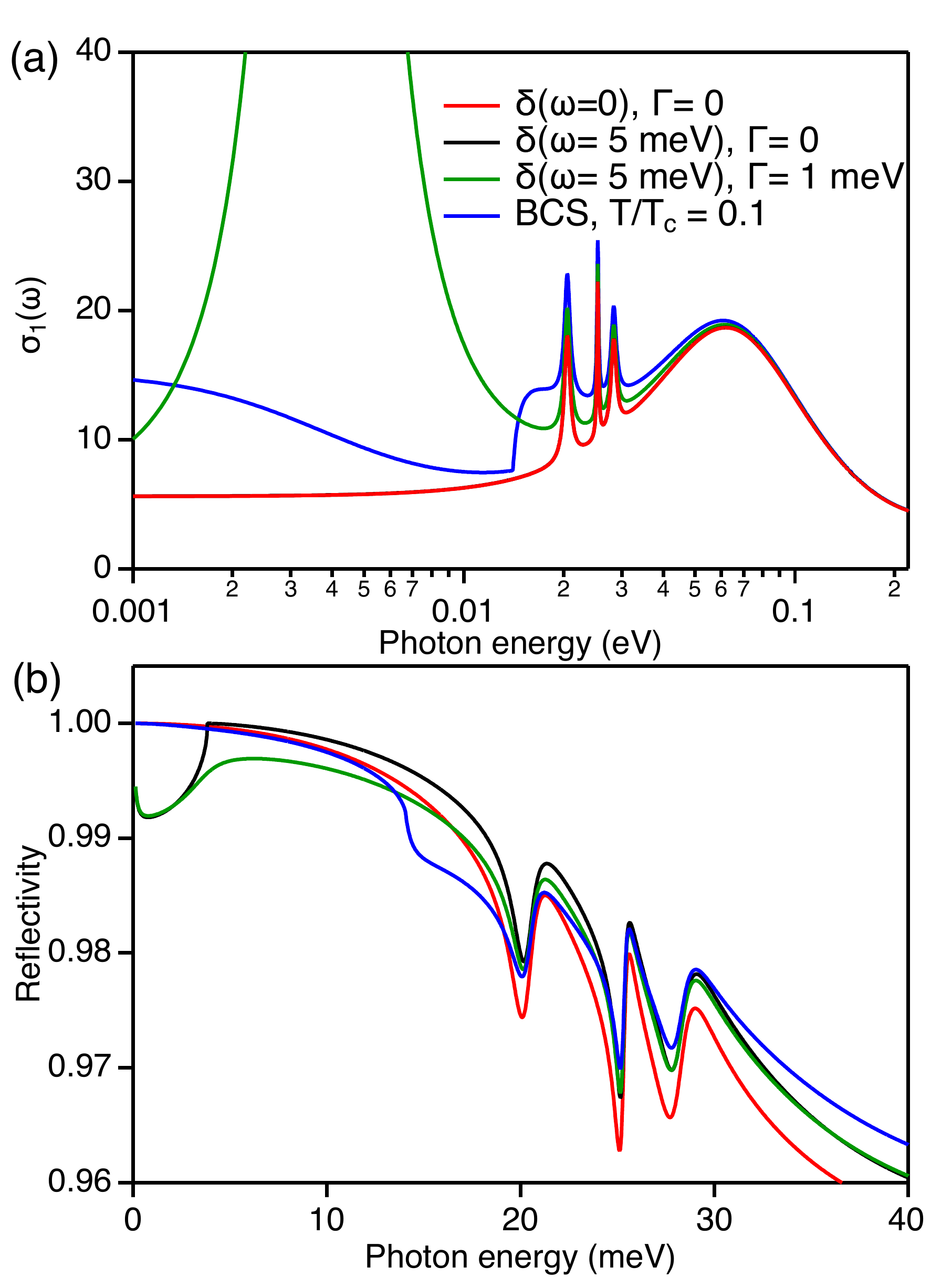}
	\caption{(Colour online) (a): Simulated optical conductivities with three possible scenarios for the distribution of low energy spectral weight between a Drude response and a collective mode (see text for detailed discussion). Also shown is a calculation using the BCS optical conductivity. Note that the $\sigma(\omega)_{1}(\omega)$ for a free (red) and pinned (black) collective mode are identical. The difference is only visible in $\sigma_{2}(\omega)$. (b): corresponding reflectivity curves. The key difference we aim to highlight is the deviation of the reflectivity from unity.}
	\label{fig:sim}
\end{figure}
The reflectivity of a fully gapped s-wave superconductor at zero temperature is unity below the superconducting gap. At finite temperature, thermal excitations result in a simultaneous normal state and superconducting response and this lowers the reflectivity. This behaviour is often modelled using a Drude\,-\,Lorentz model consisting of a zero frequency $\delta$-function contribution, a Drude function and a Lorentz oscillator to mimic interband transitions across the superconducting gap. A similar model holds for the optical response of a CDW system. 

Fig. \ref{fig:sim} shows this model (red curves) calculated with the parameters presented in Table \ref{tab:table3}. The reflectivity (panel b) is very close to unity, but continuously deviates from unity for any finite frequency. As this model is an oversimplified picture of the true optical response of an s-wave superconductor, we also show the response calculated using the numerical model of Ref. \cite{zimmermann:physC1991} (implemented in the software package RefFit \cite{Kuzmenko2005}). For this model (blue curves), we used the same Drude response as before, but replace the $\delta$-function contribution with a condensate response with the same spectral weight and calculated at a temperature corresponding to the 16 K data (i.e. T/T$_{c}\approx$\,0.1). Although the difference in the optical conductivity is significant, the changes in the reflectivity are rather modest. The most prominent feature is the distinct step that happens at the gap edge (where we have chosen $2\Delta\,=\,14 meV$). Importantly, at these elevated temperatures, one would still not expect the reflectivity to be exactly unity inside our experimental window. 

Next we show the conductivity and reflectivity for a pinned, but undamped collective mode ($\delta(\omega$\,=\,5\,meV), black curves) in addition to the Drude response . The associated optical conductivity is indistinguishable from the red curve, but the reflectivity shows a distinct difference compared to the previous models. Above the collective mode, the reflectivity increases to unity as a result of the singular response in the dielectric function associated with the collective mode. As a consequence of this a rest-strahlen type band appears, and we believe this is what we observe in our experiments. 

Finally, we also show calculations for the same model, but now with a damped collective mode response (green curves), where we have chosen a damping of 2 meV. The optical conductivity of this model would be consistent with our optical conductivity data and spectral weight analysis, but the calculated reflectivity is distinctly lower compared to our experimental data at 16 K. 

We conclude that our experimental data is most consistent with a pinned but weakly damped collective mode. Within our experimental uncertainty we can exclude a damped collective mode with width larger than 1 meV, as this lead to discrepancy with our measured reflectivity. However, we cannot exclude that the collective mode is unpinned. The real part of the optical conductivity (blue curve, Fig. \ref{fig:sim}a) right above the gap edge in our simulation changes due to the piling up of spectral weight related to optical transitions across the gap. Comparing this to the data presented in Fig. \ref{fig:s1}, we note that there is no enhancement of spectral weight visible in the data around 11\,meV. In Fig. \ref{fig:s1} we see a strong enhancement of spectral weight in the energy range above 40 meV and this would suggest that $2\Delta\approx$\,40\,-\,50 meV, much larger than what one would guess based on the reflectivity data.

Pinned or not, our data strongly indicates the presence of a sliding mode contribution. These collective modes have previously mostly been observed in quasi one-dimensional materials, such as NbSe$_3$ \cite{Ong1977}, TaS$_3$ \cite{Gruner1981}, K$_{0.3}$MoO$_3$ \cite{Degiorgi1993},  (TaSe$_4$)$_2$I \cite{Degiorgi1991}, and (TMTSF)$_2$PF$_6$ \cite{Degiorgi1991, Donovan1994}. The observation of these collective excitations is intrinsically more difficult in higher dimensional materials due to the large contribution from carriers that do not contribute to the CDW phase formation (as is the case here), but pinned modes have been observed in various transition metal oxides and Bechgaard salts \cite{kida2002, dressel2010}.  However, in TMDC materials there are very few reports, except perhaps some indirect indications of their existence (for example, Ref.'s \cite{Vaskivskyi2016, Ma2017, Liu2018, Wen2020}). One particular exception is work by Barker et al., who inferred the existence of a CDW sliding mode in 2H-TaSe$_2$ and 1T-TaS$_2$ from a spectral weight transfer analysis \cite{Barker1975}. However, more precise measurements with far-infrared data down to 3\,meV seem to rule out this interpretation \cite{Vescoli1998, Dordevic2003}. The indirect observation of the sliding mode in VSe$_{2}$ that we report here is therefore unusual and we speculate that this observation is made possible by the freezing out of normal charge carriers. This makes VSe$_{2}$ an interesting subject for future sub-THz experiments to further explore the dynamics of collective excitations in two dimensional materials. 

\begin{acknowledgments}
We thank A. de Visser for support with transport measurements, as well as H. Ellermeijer and J. Mozes for continuous technical support. Funding: EDC and CJS acknowledge funding and support from the EPSRC Centre for Doctoral Training in Condensed Matter Physics (CDT-CMP) Grant No. EP/L015544/1. 
\end{acknowledgments}

\bibliography{vse2.bib}

\newpage

\appendix

\section{Additional characterisation information.}\label{app:hr}
\begin{figure}[htb]
	\includegraphics[width=0.6\columnwidth]{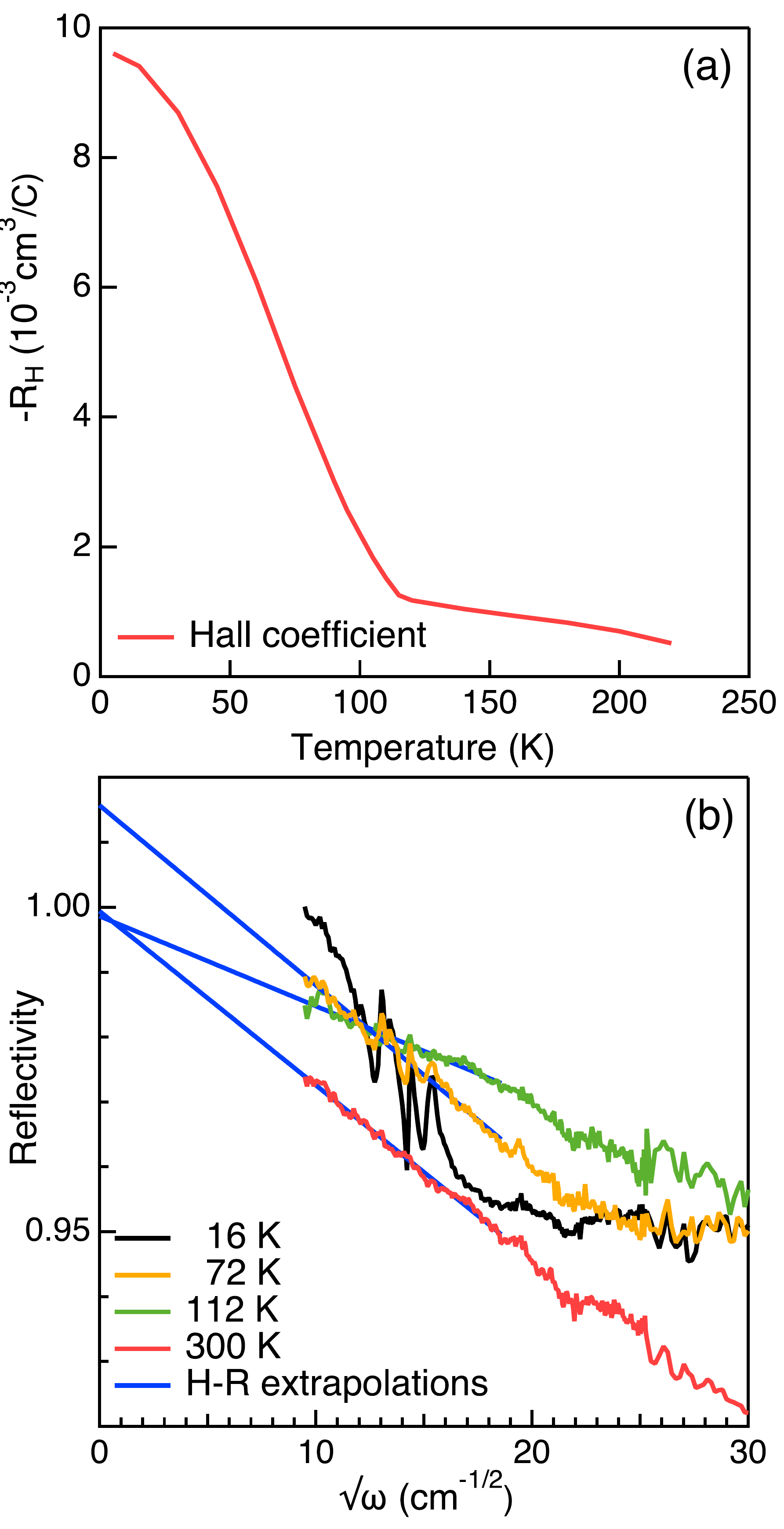}
	\caption{(Colour online) (a): Hall coefficient of a single crystal from the same batch as the crystal used for this study. (b):Reflectivity data at selected temperatures plotted as function of $\sqrt{\omega}$ together with extrapolations obtained from fits using Eq.\,\ref{eq:hr}. The reflectivity at 16\,K deviates from the Hagen-Rubens behaviour as can be seen by the clear curvature in the entire energy range.}
	\label{fig:hall}
\end{figure}
An extensive transport and x-ray photoelectron spectroscopy characterisation of the crystals used in this study has been reported in Ref. \cite{Sayers2020}. In Fig. \ref{fig:hall}a we present additional measurements of the Hall coefficient, which allows us to estimate the charge carrier contribution and the associated changes as function of temperature. Our data is overall in agreement with previously published results \cite{VanBruggen1976, Bayard1976}. We note that the Hall coefficient decreases significantly more as temperature decreases to a value R$_{H}$\,=\,-9.6$\cdot$10$^{-3}$ cm$^{3}$/C at 5\,K, compared to the previously reported R$_{H}$\,=\,-3.1$\cdot$10$^{-3}$ cm$^{3}$/C \cite{VanBruggen1976} and R$_{H}$\,=\,-2$\cdot$10$^{-3}$ cm$^{3}$/C \cite{Bayard1976}. This may attest further to the high quality of the crystals used in this study.
 
The Hagen-Rubens relation is used to examine the low frequency behaviour of free electrons. It is derived from the Drude model by making the approximation $\omega\leq 1/\tau$. This gives (in CGS units):
\begin{eqnarray}\label{eq:hr}
\left. R(\omega)\approx 1-\sqrt{\frac{2\omega}{\pi\sigma_0}}~~(\omega\ll\gamma)\right.
\end{eqnarray}
in which $\gamma$ is the scattering rate of the Drude response, and $\sigma_0$ is the DC conductivity. Even though VSe$_2$ is metallic in both the normal and CDW phase, the low frequency electronic responses of these two phases is clearly different. The reflectivity data above $T_{c}$ agrees well with the Hagen-Rubens relation, which is linear with respect to $\sqrt{\omega}$ and extrapolates to unity at zero frequency (see Fig. \ref{fig:hall}b). Below the critical temperature, the data starts to deviate from the Hagen-Rubens behaviour in the sense that the extrapolation to zero frequency becomes larger than unity (see for example the 72\,K extrapolation) and at even lower temperatures develops curvature down to the lowest measured frequency due to an interplay between the remaining Drude response and the emerging collective mode and low energy optical transition.

\section{Drude-Lorentz models and fits to the reflectivity}\label{app:dl}
\begin{figure}[htb]
	\includegraphics[width=0.95\columnwidth]{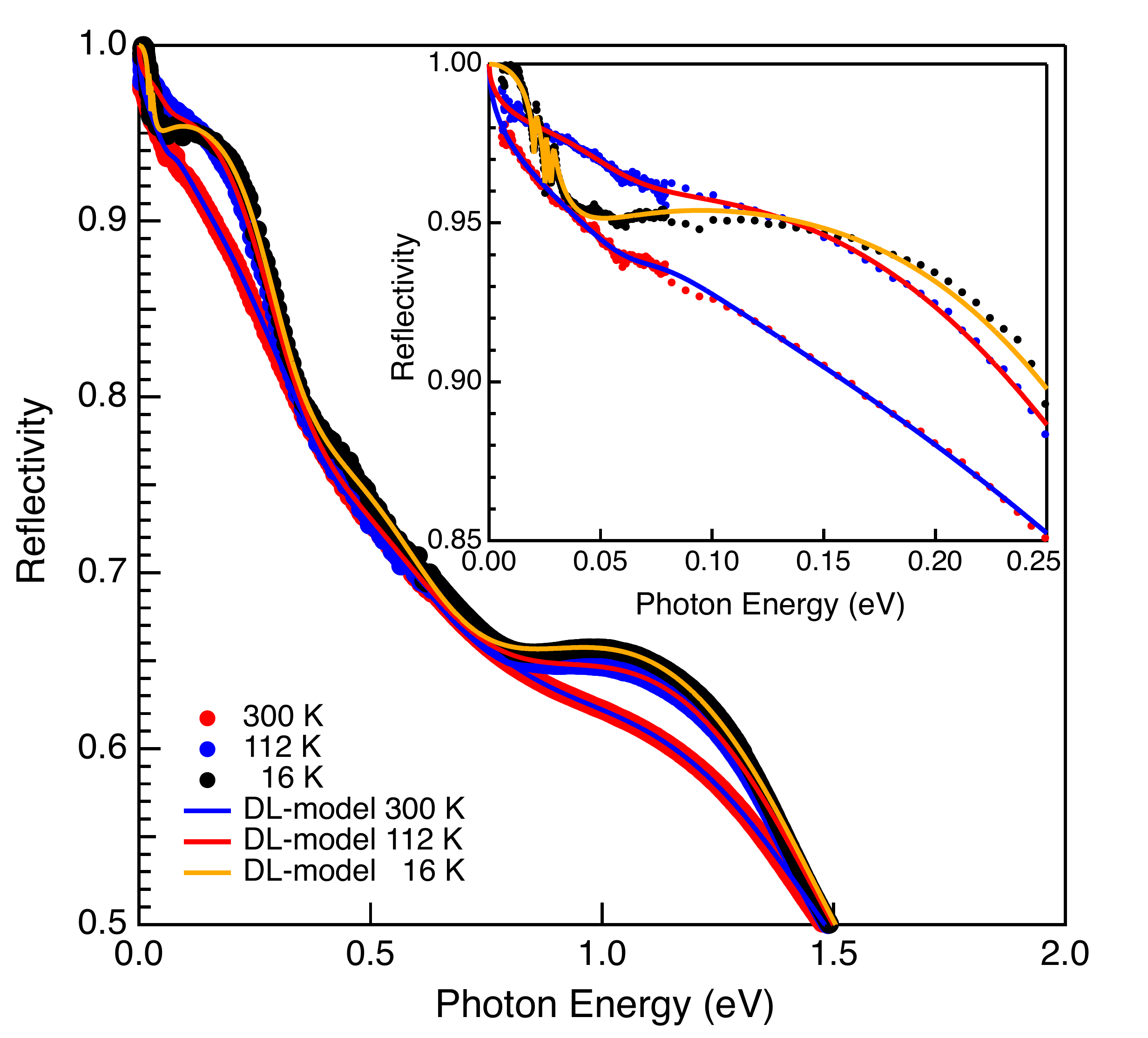}
	\caption{(Colour online) Experimental reflectivity compared to the Drude-Lorentz model fits calculated from the parameters listed in Table \ref{tab:table3} using Eq. \ref{DL_DF}. The inset shows the low energy fits.}
	\label{fig:fits}
\end{figure}
Before using the variational dielectric function method \cite{Kuzmenko2005} to extract optical functions from reflectivity data, we use a series of Drude-Lorentz models to fit the reflectivity at all measured temperatures. These models are given by,
\begin{eqnarray}\label{DL_DF}
\hat{\varepsilon}(\omega)=1-\frac{\omega_{p}^{2}}{\omega(\omega+i\Gamma)}-\sum_{i}\frac{4\pi f_{i}^{2}}{i\omega\Gamma_{i}-(\Omega_{0,i}^{2}-\omega^{2})}\nonumber\\
-\sum_{ph}\frac{4\pi f_{ph}^{2}}{i\omega\Gamma_{ph}-(\Omega_{0,ph}^{2}-\omega^{2})}
\end{eqnarray}
The first term corresponds to a Drude term, while the sum over $i$ indicates interband transition and the sum over $ph$ corresponds to phonon modes. The parameters appearing here are the plasma frequency and scattering rate ($\omega_{p}$ and $\Gamma$), oscillator strengths ($f_{i}$ and $f_{ph}$), resonance frequencies ($\Omega_{0,i}$ and $\Omega_{0,ph}$) and widths ($\Gamma_{i}$ and $\Gamma_{ph}$). We use this model to make a best fit to our optical data at all measured temperatures. Here we present the 16\,K, 112\,K and 300\,K data, fits (see Fig. \ref{fig:fits}) and models (table \ref{tab:table3}). Note that the 16\,K data only has one Drude component and instead a $\delta(\omega)$-function contribution representing the collective mode as discussed extensively in the main text. The Drude-Lorentz model presented here is used to extend the reflectivity data outside our measured range. Parameters are optimized to describe the best fit to data at a particular temperature as demonstrated in Fig. \ref{fig:fits}. The low energy parameters show a non-monotonous temperature variation, reflecting to some degree the spectral weight transfer associated with the formation of the CDW phase. The high energy Lorentz parameters are fairly temperature independent and can serve as accurate benchmark for the energy positon of interband transitions.
\begin{table*}[htb]
\caption{\label{tab:table3}Drude-Lorentz models.\\ The first two terms are Drude terms with plasma frequency, $\omega_{p}$ and scattering rate $\gamma_{D}$. At low temperature, we need a $\delta$-function contribution that we have indicted by "collective mode'. Terms labeled with Lorentz ${i}$ describe interband terms and have parameters $\omega_{0}$, $f_p$ and $\gamma$, (eigenfrequency, oscillator strength and scattering rate).  In addition, we find $\varepsilon_\infty=23.6$. The last rows summarize the phonon parameters at the lowest measured temperature. All parameters presented are in meV.}
\resizebox{0.8\textwidth}{!}{
\begin{ruledtabular}
\begin{tabular}{cccccc}
Temperature&&300\,K&112\,K&16\,K&\\
\hline
\\
\multirow{2}{*}{Drude}
&$\omega_p$&3795.2&2168.0&1505.2\\
&$\gamma_D$&203.3&313.7&57.0\\
\\
\hline
\\
\multirow{2}{*}{Drude}
&$\omega_p$&1203.0&2830.6&-\\
&$\gamma_D$&28.3&30.4&-\\
\\
\hline
\\
\multirow{1}{*}{Collective mode}&$A$&-&-&1569.6\\
\\
\hline
\\
\multirow{3}{*}{Lorentz 1}&$\omega_{0}$&69.7&79.4&66.1\\
&$f_p$&536.0&1973.2&3340.2\\
&$\gamma$&32.2&97.3&97.2\\
\\
\hline
\\
\multirow{3}{*}{Lorentz 2}&$\omega_{0}$&600.7&514.5&471.1\\
&$f_p$&5002.8&4021.4&3685.5\\
&$\gamma$&1026.0&685.0&534.4\\
\\
\hline
\\
\multirow{3}{*}{Lorentz 3}&$\omega_{0}$&1143.8&1068.9&1037.1\\
&$f_p$&4400.2&5186.3&5530.3\\
&$\gamma$&915.0&821.4&814.6\\
\\
\hline
\\
\multirow{3}{*}{Lorentz 4}&$\omega_{0}$&2003.6&2003.6&2003.6\\
&$f_p$&2354.5&2575.2&2662.6 \\
&$\gamma$&943.5&943.5&943.5\\
\\
\hline
\\
\multirow{3}{*}{Lorentz 5}&$\omega_{0}$&2564.6&2564.6&2564.6\\
&$f_p$&3208.1&3208.1&3208.1\\
&$\gamma$&1262.8&1262.8&1262.8\\
\\
\hline
\\
\multirow{3}{*}{Phonon 1}&$\omega_{0}$&-&-&20.5\\
&$f_p$&&&256.0\\
&$\gamma$&&&1.0\\
\\
\hline
\\
\multirow{3}{*}{Phonon 2}&$\omega_{0}$&-&-&25.4\\
&$f_p$&&&166.1\\
&$\gamma$&&&0.4\\
\\
\hline
\\
\multirow{3}{*}{Phonon 3}&$\omega_{0}$&-&-&28.5\\
&$f_p$&&&239.9\\
&$\gamma$&&&1.2\\
\end{tabular}
\end{ruledtabular}
}
\end{table*}

\section{Consistency of the optical data with transport experiments.}
\begin{figure}[htb]
	\includegraphics[width=0.9\columnwidth]{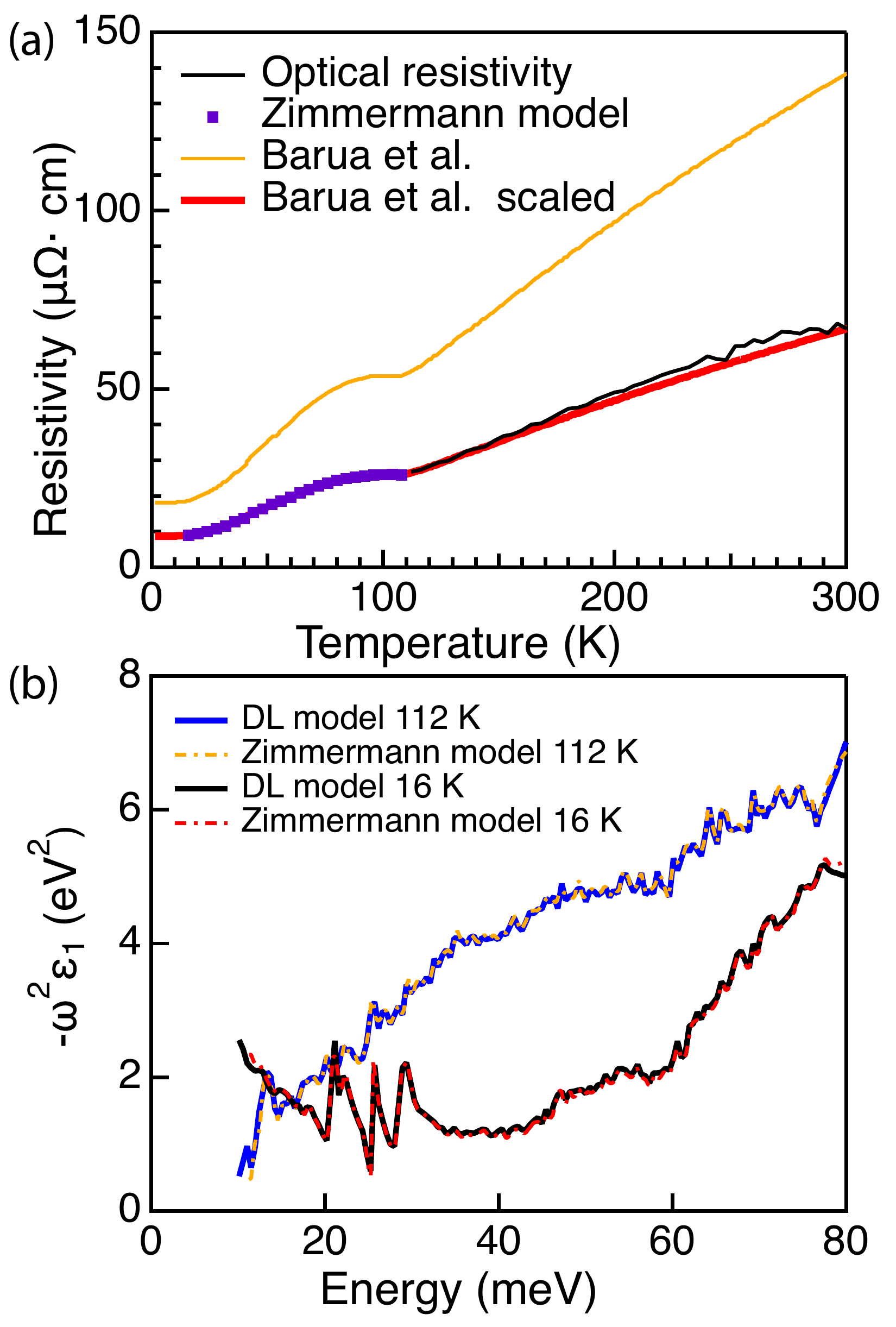}
	\caption{(Colour online) (a): comparison between resistivity data from Ref. \cite{Barua:SRep2017} and extrapolated conductivity data. For this comparison the data from Ref. \cite{Barua:SRep2017} is scaled to match room temperature DC optical data. At low temperature, open symbols indicate the extrapolation obtainedd from the BCS-type model from Ref. \cite{zimmermann:physC1991}. (b): comparison between the estimate of the sliding mode contribution (see Fig. \ref{fig:E1} and discussion) with and without broadening of the $\delta(\omega)$ response.}
	\label{fig:rho}
\end{figure}
From the Hagen-Rubens analysis or the determination of the optical conductivity, we can estimate the DC conductivity, or alternatively the DC resistivity. To test the reliability of our optical data, we compare these extrapolations to previously published results. Different reports of transport experiments show some degree of variation in the absolute values, but the overall temperature dependence is very similar \cite{Bayard1976, VanBruggen1976, Pasztor2017, Barua:SRep2017, Sayers2020}. This is perhaps due to small differences in crystal quality, but another likely factor is the uncertainty that is posed by estimating thicknesses and contact distances on these small, thin crystals. To test the agreement of our optical conductivity data with transport experiments, we thus choose one previously reported result, Ref. \cite{Barua:SRep2017} and scale their transport data to match with our extrapolated DC resistivity data at room temperature. Fig. \ref{fig:rho} shows that in the normal state we obtain excellent agreement between our conductivity data and transport data  

As explained in the main text, we find that the low temperature conductivity has a sliding mode contribution. Our Drude-Loretnz models include this contribution as a $\delta(\omega)$-function contribution (see table \ref{tab:table3}). Such a description is however only accurate at zero temperature, while at finite temperature it gives the wrong zero frequency extrapolation for the DC conductivity. At any finite temperature, excitations of the sliding mode will broaden the $\delta(\omega)$-function response. The BCS type model by Zimmermann \cite{zimmermann:physC1991}, discussed in the main text, allows us to estimate this finite temperature response more accurately. 
The outcome is shown in Fig. \ref{fig:rho}a as solid symbols. Using the correct temperature and spectral weight for the sliding mode, we obtain excellent agreement with the low temperature transport data, provided that we allow a small additional broadening of (less than 0.125 meV) of the collective mode. Such impurity broadening is expected to occur in imperfect crystals or for small domains of CDW order. Importantly, the inclusion of impurity broadening does not change the finite frequency response in our measurement window. As an example, we show the comparison of the sliding mode spectral weight as obtained from the dielectric function with and without sliding mode broadening (see Fig. \ref{fig:E1} and the discussion pertaining to that figure). This comparison shows that for sufficiently weak impurity broadening, the function $-\omega^{2}\varepsilon_{1}(\omega)$ provides a reliable means to estimate the sliding mode spectral weight. 

\section{The imaginary component of the optical conductivity}\label{app:imsig}
In the main manuscript we present a method to determine the contribution of the collective mode from the dielectric function. An equivalent method makes use of the imaginary component of the optical conductivity. These two functions are related through:
\begin{figure}[htb]
	\includegraphics[width=0.9\columnwidth]{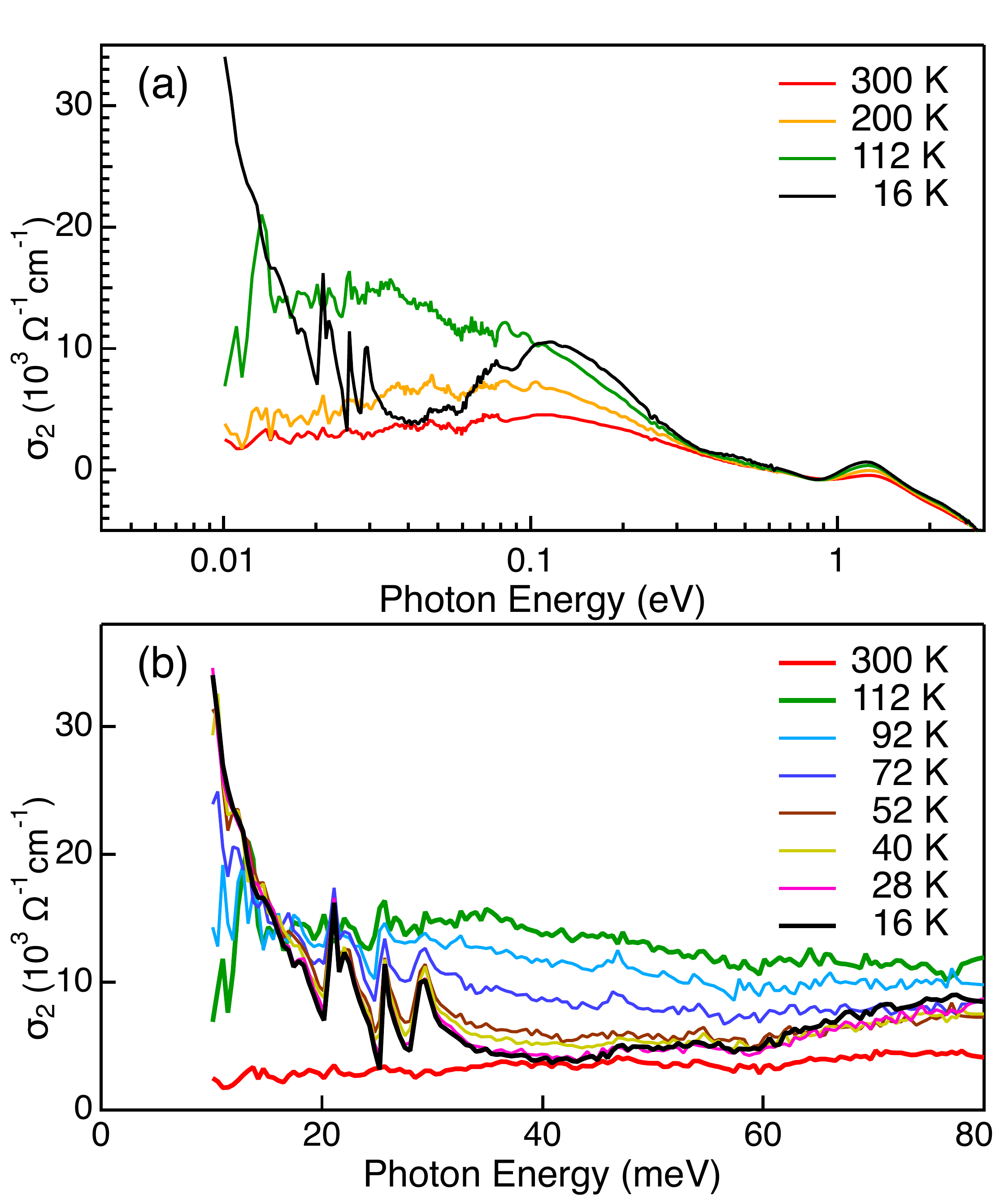}
	\caption{(Colour online) (a): Imaginary component of the optical conductivity at selected temperatures. (b) additional temperatures to illustrate the emergence of the $\omega^{-1}$ divergence associated with the collective mode.}
	\label{fig:s2}
\end{figure}
\begin{equation}\label{eq:eps2s}
\sigma_{2}(\omega)\propto-\frac{\omega\varepsilon_{1}(\omega)}{4\pi}
\end{equation} 
In the main text we show that $\varepsilon_{1}\,=\,A/\omega^{2}$. Together with Eq. \ref{eq:eps2s}, we therefore expect:
\begin{equation}
\sigma_{2}(\omega)\propto\frac{A}{\omega}
\end{equation} 
Fig. \ref{fig:s2} shows the imaginary component of the optical conductivity at the same temperatures as those presented in Fig. \ref{fig:s1} of the manuscript. Below $T_{CDW}$, we indeed see the emergence of the collective mode response as a $\omega^{-1}$ divergence.

\section{Matrix elements}\label{app.me}
Previous work has shown that the weak-coupling charge density-wave phase of VSe$_2$ generates small suppressions of spectral weight, in regions of the bandstructure close to the Fermi level that are separated by one CDW wave-vector \cite{Henke2020}. This is corroborated by the minimal change in density of states near $E_F$ as seen by scanning tunnelling microscopy \cite{Jolie2019}. Based on this, the small CDW gap that opens is expected to affect only a few percent of the Fermi surface. Since the optical conductivity is additive, the only way that the opening of such a small gap could explain a strong suppression of the free carrier response would be if the optical matrix elements (as defined in the Kubo-Greenwood formula) have a significant maximum around the regions where gaps open.

To test whether this could be the case, we constructed a tight-binding model by considering symmetry-allowed hopping (in the Slater-Koster formalism \cite{Slater_PR_1954}) between Vanadium $d_{xy}$, $d_{yz}$ and $d_{xz}$ orbitals and Selenium $p_x$, $p_y$ and $p_z$ orbitals in the $1T$-VSe$_2$ lattice, up to second nearest neighbours. This generates a $9\times9$ Hamiltonian, which is fitted to the \textit{ab initio} bandstructure given in \cite{Henke2020}. We then compute the matrix elements, given by $|\langle \Psi_{\mathbf{k}} | \nabla_{\mathbf{k}_i} \hat{H} | \Psi_{\mathbf{k}}\rangle |^2$. Here, $|\Psi_k \rangle$ is the eigenvector of the tight-binding Hamiltonian at momentum $\mathbf{k}$, and $\nabla_{\mathbf{k}_i} \hat{H}$ is the gradient of the Hamiltonian along $\mathbf{k}_i$ ($i=\{x,y,z\}$), where the latter is the momentum direction parallel to the probed direction of conductivity. Since our experiments probe the in-plane optical response, only the in-plane directions ($k_x$ and $k_y$) should be relevant.
\begin{figure}[htb]
	\includegraphics[width=1\columnwidth]{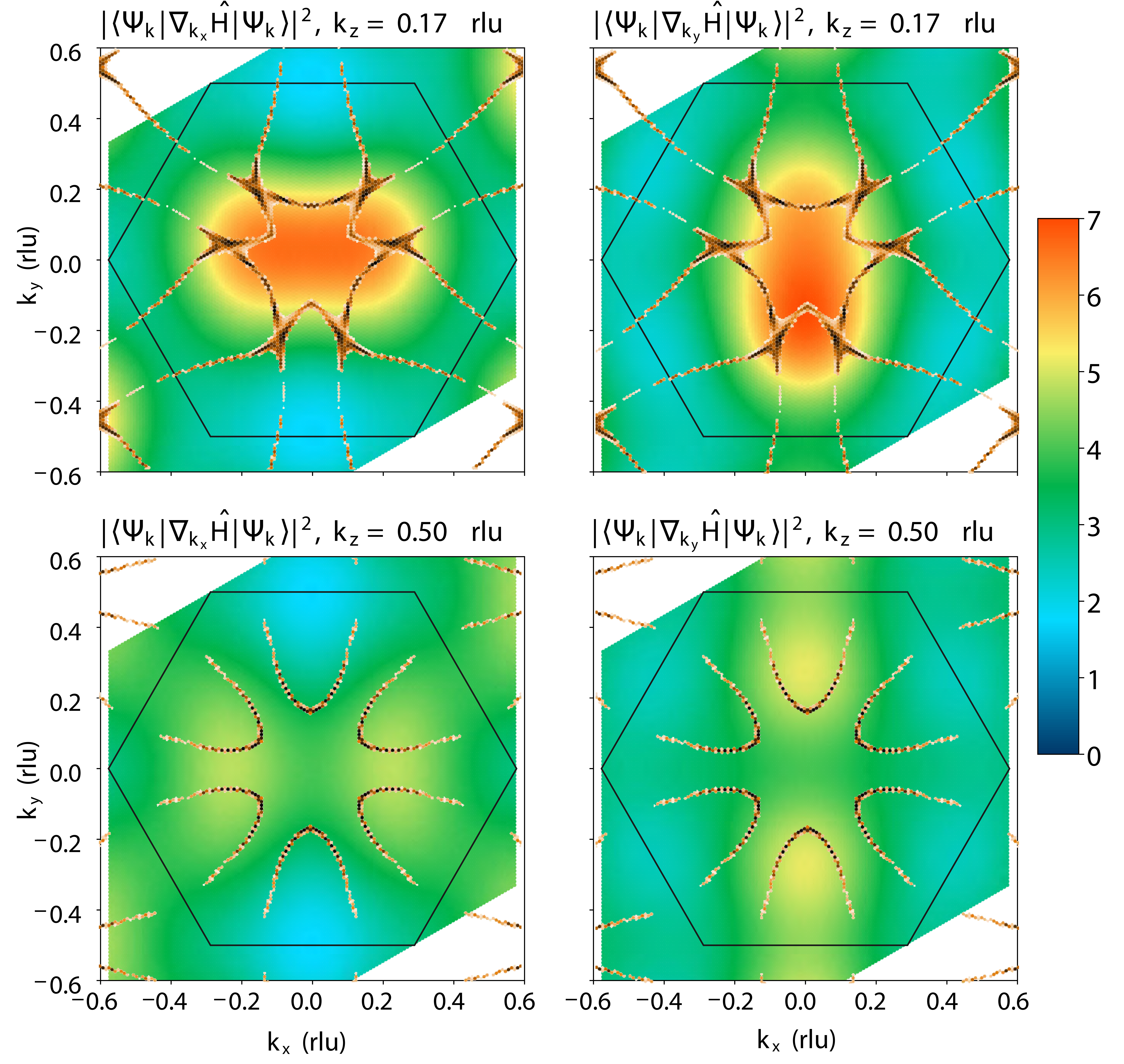}
	\caption{(Colour online) Matrix elements at selected $k_{z}$ momenta on fixed colorscale with arbitrary units. The Fermi surface contours in the CDW phase are overlaid to illustrate the sections of the Fermi surface that contribute most. }
	\label{fig:ME}
\end{figure}
As demonstrated by Figure \ref{fig:ME}, the maximum variation of the matrix elements in the $k_x$-$k_y$ plane at the two values of $k_z$ where the largest gaps are expected to open \cite{Henke2020} is less than a factor of four. Additionally, the maxima of the matrix elements lie away from the regions of the Brillouin zone where gaps are expected to open. We therefore conclude that there is no reason why the optical conductivity would be mostly sensitive to the gapped parts of the Fermi surface.

\end{document}